\documentclass[aps,prd,onecolumn, groupedaddress, showpacs]{revtex4} % 1 column

\usepackage{graphicx}

\begin{document}

\title{Induced gravitational wave background and primordial black holes}

\author{Edgar Bugaev}
\email[e-mail: ]{bugaev@pcbai10.inr.ruhep.ru}
%\homepage[]{Your web page}
%\thanks{}
%\altaffiliation{}

\author{Peter Klimai}
\email[e-mail: ]{pklimai@gmail.com}
%\homepage[]{Your web page}
%\thanks{}
%\altaffiliation{}
\affiliation{Institute for Nuclear Research, Russian Academy of
Sciences, 60th October Anniversary Prospect 7a, 117312 Moscow,
Russia}

%\author{Edgar Bugaev and Peter Klimai}

%\address{ \bigskip
%Institute for Nuclear Research, Russian Academy of Sciences,
%\\ 60th October Anniversary Prospect 7a, 117312 Moscow, Russia }

%\date{\today}

\begin{abstract}

We calculate the frequency dependence of gravitational wave background arising at second order
of cosmological perturbation theory due to mixing of tensor and scalar modes. The calculation
of the induced gravitational background is performed for two special cases: for the power
spectrum of scalar perturbations which has a peak at some scale and for the scalar spectrum
predicted by the inflationary model with the running mass potential. We show that the amplitudes
of the induced gravitational background, in the frequency region $\sim 10^{-3} - 10^3\;$Hz,
are effectively constrained by results of studies of primordial black hole production in early universe.
We argue that though today's LIGO bound on ${\cal P}_{\cal R}(k)$ is weaker than the PBH one,
Advanced LIGO will be able to set a stronger bound, and in future the ground-based
interferometers of LIGO type will be suitable for obtaining constraints on PBH number
density in the mass range $\sim 10^{11} - 10^{15}$ g.

\end{abstract}

 %%% \hfill arXiv:XXXX.XXXX [astro-ph]}
% 98.80.–k = Cosmology
% 04.70.-s = Physics of black holes
% 04.30.Db = Gravitational waves: theory / Wave generation and sources

\maketitle

\section{Introduction}
\label{sec-intro}

According to the one of the main predictions of General Relativity, the present Universe is
filled with a diffuse gravitational background produced by sources of astrophysical
and cosmological origin (see reviews \cite{Allen:1996vm,Maggiore:1998jh,Buonanno:2003th}).
In particular, relic stochastic background
necessarily arises in all standard and superstring-motivated
(``pre-big bang'' (PBB) \cite{Gasperini:2002bn})
inflationary scenarios,
as a result of the process of the amplification of vacuum fluctuations (such an amplification
of quantum fluctuations of the gravitational field modes had been discovered in
\cite{Grishchuk:1974ny} and had been used in a model with de Sitter phase of expansion
in \cite{Starobinsky:1979ty}). The standard inflationary
scenario is characterized by a flat or slightly decreasing spectrum
\cite{Rubakov:1982df, Fabbri:1983us, Abbott:1984fp, Starobinsky:1985ww}, which is constrained
at the scale of the present Hubble radius (at $f\sim 10^{-18}$ Hz) by the large scale isotropy
of the CMB radiation \cite{Krauss:1992ke, Polarski:1995jg} .
On the contrary, PBB models predict a growing spectrum \cite{Gasperini:1992em, Gasperini:1992dp}
constrained at high frequencies by the nucleosynthesis bound \cite{Schwarztmann, Brustein:1996ut}.

In the period near the end of inflation and after it the processes are possible which could
result in additional gravitational background: in particular, first order phase transitions
\cite{Turner:1990rc, Kosowsky:1991ua} (such
transitions can occur even before the end of inflation, like, e.g., in some two-field
inflationary models \cite{Copeland:1994vg}) and preheating (see, e.g.,
\cite{Khlebnikov:1997di, Dufaux:2007pt}). These processes lead to
gravitational wave radiation due to local strong inhomogeneities generated in the cosmological
fluid during the transition from inflationary expansion to radiation era and
subsequent phase transitions. The corresponding spectra of the background radiation
have a peak at some frequency (the peak's position depends on the inflationary
energy scale  and on the temperature of the phase transition), see, e.g.,
\cite{Grojean:2006bp, Easther:2006vd}.

The rather substantial gravitational wave background (GWB) is possible in cosmic string
scenarios (see, e.g., the recent work \cite{DePies:2007bm}), in brane world models
(see, e.g., \cite{Hogan:2000is}),
in scenarios with intense production of primordial black holes (PBHs) in early universe
(in the latter case GWB appears as a result of PBH evaporations \cite{BisnovatyiKogan:2004bk,
Anantua:2008am}).

It was realized during last decade that there is still another background of stochastic GWs of
cosmological origin. Namely, GWB arises as a result of non-linear evolution (i.e., of gravitational
instability) of curvature (density) fluctuations. The nonlinear effects (the mixing of
different modes) appear already in second-order of cosmological perturbation theory
\cite{Matarrese:1993zf, Matarrese:1997ay, Carbone:2004iv}.
In particular,
a second order contribution to the tensor mode, $h_{ij}^{(2)}$, depends quadratically on the first order
scalar metric perturbation, i.e., the observed scalar spectrum sources the generation of
secondary tensor modes. By other words, the stochastic spectrum of second order GWs is induced
by the first order scalar perturbations. Calculations of $\Omega_{GW}$ at second order and
discussions on perspectives of measurements of the second order GWs are contained in works
\cite{Mollerach:2003nq, Ananda:2006af, Baumann:2007zm, Saito:2008jc}.

The generation of GWs from primordial density perturbations on very small scales which are not directly
studied by astronomical measurements could be used for constraining overdensities on these
scales, in a close analogy with the case of primordial black holes \cite{Ananda:2006af,
Assadullahi:2009jc}. Large curvature perturbations leading to ``features'' (e.g., peaks
or spikes) in the primordial power spectrum and to possible PBH production, can arise in multiple
field scenarios at the end of inflation (during the preheating era) or between two consecutive
stages of inflation, as a result of parametric resonance or tachyonic instability. Such features
can, in principle, exist even in single-field inflationary models (see, e.g., \cite{Bugaev:2008bi}).
These peculiarities of the primordial scalar spectrum lead to local enhancements in the induced
spectrum of gravitational perturbations \cite{Ananda:2006af, Saito:2008jc}. Another example of
an inflationary model predicting large amplitudes of the density perturbations at small scales
is the inflationary model with the running mass potential. In a case of the positive running, the
scalar spectrum at large values of $k$ (near $k_{end}$) can steeply rise with $k$ (which is also
a kind of the ``feature'').

The aim of the present paper is two-fold: we calculate a spectrum of induced GWs for two cases,
which are not studied in previous works \cite{Ananda:2006af, Baumann:2007zm, Saito:2008jc}:
for a scalar power spectrum with a peak of non-zero width and for a scalar power spectrum
with the running of the spectral index (the latter case is studied with using the particular
model, namely, a model with the running mass potential \cite{Stewart:1996ey, Stewart:1997wg}).
The second aim of the paper is a constraining of the induced GWB using the results of PBH searches.
We consider, in the present paper, the tensor spectrum in a rather narrow interval of wave numbers,
corresponding to modes leaving horizon at the time near the end of inflation. These modes enter
the Hubble scale during the radiation-dominated (RD) era. Overdensities lead to production
of PBHs with small masses ($\sim 10^{11} - 10^{15}$ g) which have enough of time for evaporation
during the life of the universe.
Products of evaporation of these PBHs contribute to extragalactic diffuse photon and neutrino
backgrounds (which are measured experimentally). This allows to obtain constraints on the primordial
power spectrum amplitudes. An independent constraint on the scalar spectrum can, in principle,
be obtained by a direct detection of induced GWs \cite{Saito:2008jc, Assadullahi:2009jc}.
In this paper we compare the abilities of two methods of such a constraining.

The plan of the paper is as follows. In Sec. \ref{sec-2} calculation of the induced GW background
is performed for two different cases: for the delta-function power spectrum of the
primordial scalar perturbations and for the power spectrum with a peak of finite width.
The comparison of PBH constraints for the induced GWB with possible constraints
from the future experiments, such as Advanced LIGO, is given. In Sec. \ref{sec-RM} the analogous calculation
and comparison are done for the scalar spectrum predicted by the running mass model. The last
section contains our summary and conclusions.

\section{GW background calculation}
\label{sec-2}

\subsection{Connection between frequency and horizon mass}

For a wave with comoving wave number $k$ and wavelength $\lambda=2\pi/k$, propagating at the speed of light
$c$, the corresponding frequency is $f=c/\lambda $, or
\begin{equation}
f=\frac{ck}{2\pi} = 1.54\times 10^{-15 } \left( \frac{k}{{\rm Mpc}^{-1}} \right) {\rm Hz}.
\label{fck}
\end{equation}
From the constancy of the entropy in the comoving volume, we have the relation between the scale factor $a$,
temperature $T$ and the effective number of degrees of freedom $g_*$:
\begin{equation}
a\sim g_*^{-1/3} T^{-1}.
\end{equation}
From the Friedmann equation ($H^2 \sim \rho$), we have
\begin{equation}
H \sim a^{-2} g_*^{-1/6}, \label{Hag}
\end{equation}
and the horizon mass corresponding to the scale factor $a$ evolves during the radiation-dominated (RD)
epoch as
\begin{equation}
M_h \sim (H^{-1})^3 \rho \sim a^2 g_*^{1/6}. \label{Mhag}
\end{equation}
From (\ref{Hag}) and (\ref{Mhag}), the wave number of the mode entering horizon at the moment of
time $t$ (at this time, $k=aH$) is related to the horizon mass at the same moment of time by
\begin{equation}
k = k_{eq} \left( \frac{M_h}{M_{eq}} \right)^{-1/2} \left( \frac{g_*}{g_{* eq}} \right)^{-1/12}
\approx 2 \times 10^{23} (M_h[{\rm g}])^{-1/2} \;\; {\rm Mpc}^{-1},
\label{kkeq}
\end{equation}
where in the last equality we have adopted that $g_{* eq} \approx 3$, $g_* \approx 100$,
\begin{equation}
M_{eq} = 1.3 \times 10^{49} {\rm g} \cdot (\Omega_m h^2)^{-2} \approx 8 \times 10^{50} {\rm g},
\end{equation}
\begin{equation} \label{keq}
k_{eq} = a_{eq} H_{eq} =  \sqrt{2} H_0 \Omega_m \Omega_R^{-1/2} \approx 0.0095 \; {\rm Mpc}^{-1}.
\end{equation}

The frequency of the wave corresponding to the wave number $k$ can be related to the horizon mass
by the relation following from (\ref{fck}) and (\ref{kkeq}),
\begin{equation}
f \approx 3 \times 10^8 \;{\rm Hz} \times (M_h[{\rm g}])^{-1/2} ; \;\;\;
M_h \approx \frac{9 \times 10^{16} {\rm g}}{(f[{\rm Hz}])^2}.
\label{fMh}
\end{equation}
For scalar-induced GWs, the single mode in scalar spectrum does not correspond to the only
one mode in ${\cal P}_h$. For example, for the $\delta$-function-like spectrum
${\cal P}_{\cal R}(k) \sim \delta(k-k_0)$, the GW spectrum is continuous and stretches from $0$
to $2k_0$ \cite{Ananda:2006af}. However, the order of magnitude of wave numbers of induced GWs, as we will see,
is the same as of scalar perturbations, so (\ref{fMh}) gives an estimate of GW frequency
that will be generated from perturbations entering horizon at its mass scale $M_h$. Furthermore, if
PBHs form from a scalar spectrum of perturbations at a horizon mass scale $M_h$, the typical
PBH mass will be of order of $M_h$ (see, e.g., \cite{Bugaev:2008gw}), so (\ref{fMh})
relates the typical PBH mass with the characteristic frequency of second-order GWs produced.

\subsection{General formulas}

The components of the spatially flat FRW metric are following:
\begin{eqnarray}
g_{00} & = & - a^2 (1+ 2 \Phi ), \\
g_{0i}& = &a^2 \left(\partial_i \omega + \omega_i \right), \\
g_{ij}& = & a^2 \left[ \left( 1 - 2 \Psi \right) \delta_{ij} +
D_{ij} h + \partial_i h_j + \partial_j h_i + \frac{1}{2} h_{ij} \right].
\end{eqnarray}
For a derivation of the equation of motion for second order tensor perturbations we used
the generalized longitudinal gauge, which is defined by the relations
\begin{equation}
\omega^{(1)} = h^{(1)} = h_i^{(1)} = 0.
\end{equation}
Besides, we neglect, following \cite{Baumann:2007zm}, the first order vector and tensor perturbations,
\begin{equation}
\omega_i^{(1)} =  h_{ij}^{(1)} = 0.
\end{equation}
With these approximations the components of the second order Einstein tensor, ${G^i_j}^{(2)}$, depend
only on first-order perturbations $\Phi^{(1)}$, $\Psi^{(1)}$ and second-order perturbations
$\Phi^{(2)}$, $\Psi^{(2)}$, $\omega^{(2)}$, $\omega_i^{(2)}$, $h^{(2)}$, $h_i^{(2)}$ and
$h_{ij}^{(2)}$. The second-order energy-momentum tensor, with ignoring
anisotropic stress contributions (in this case $\Phi=\Psi$), is given by
\begin{equation}
{T^i_j}^{(2)} = \frac{1}{2} w \delta^{(2)}(\rho) \delta^i_j + \rho^{(0)}(1+w) v^i_{(1)} v_j^{(1)}.
\end{equation}
Here, $v^i_{(1)}$ is the vector (solenoidal) part of the first-order velocity perturbation,
\begin{equation}
v_i^{(1)} = - \frac{2}{3 {\cal H}^2 (1 + w)} \partial_i
 \left( {\Psi^{(1)}}' + {\cal H} \Psi^{(1)}  \right),
\end{equation}
where ${\cal H} = aH$ and $w=p^{(0)}/\rho^{(0)}$ is the equation of state parameter. Using the
Einstein equations, with keeping in them only transverse-traceless contributions,
\begin{equation} \label{TT}
\left[ {G_i^j}^{(2)} \right]^{TT} = 8 \pi G \left[ {T_i^j}^{(2)} \right]^{TT},
\end{equation}
leads to the evolution equation for the second-order tensor perturbation $h_{ij}$, which contains,
in the source term, only first-order scalar perturbations (all the second-order perturbations
except of $h_{ij}^{(2)}$ are eliminated):
\begin{equation}
h_{ij}'' + 2 {\cal H} h_{ij}' - \nabla^2 h_{ij} = -4 \left[
4 \Psi \partial_i \partial_j \Psi + 2 \partial_i \Psi \partial_j \Psi -
\frac{4}{3(1+w) {\cal H}^2} \partial_i (\Psi'+  {\cal H} \Psi) \partial_j (\Psi'+  {\cal H} \Psi)
\right]^{TT}.
\end{equation}
Here, we have omitted the indexes: $\Psi^{(1)} \to \Psi$, $h_{ij}^{(2)} \to h_{ij}$.
The elimination of all second-order perturbations except of $h_{ij}^{(2)}$ is due to using of transverse traceless
projection in Einstein equations (see Eq. (\ref{TT})). It had been
shown in \cite{Ananda:2006af, Baumann:2007zm} (and can be straightforwardly checked) that
all the second-order terms which enter the Einstein tensor
(except of $h_{ij}^{(2)}$) are canceled after such a projection.

According to \cite{Baumann:2007zm}, the power spectrum of induced GWs is given by the expression
\begin{equation} \label{Phktau}
{\cal P}_h(k, \tau) = \int\limits_0^\infty d\tilde k \int \limits_{-1}^{1} d\mu  \;
{\cal P}_\Psi(|{\bf k-\tilde k|}) {\cal P}_\Psi (\tilde k) {\cal F}(k,\tilde k,\mu, \tau),
\end{equation}
where
\begin{eqnarray}
\label{calF}
{\cal F}(k,\tilde k,\mu, \tau) = \frac{(1-\mu^2)^2}{a^2(\tau)} \frac{k^3 \tilde k^3}{|{\bf k-\tilde k}|^3}
\int \limits_{\tau_0}^{\tau} d \tilde \tau_1 \; a(\tilde \tau_1) g_k(\tau, \tilde\tau_1)
f({\bf k},{\bf \tilde k}, \tilde \tau_1) \times \nonumber \\ \times
\int \limits_{\tau_0}^{\tau} d \tilde \tau_2 \; a(\tilde \tau_2) g_k(\tau, \tilde\tau_2)
\left[f({\bf k},{\bf \tilde k}, \tilde \tau_2) + f({\bf k},{\bf k-\tilde k}, \tilde \tau_2)\right]
\end{eqnarray}
and
\begin{equation} \label{f}
f({\bf k},{\bf \tilde k}, \tau) = 12 \Psi(\tilde k\tau) \Psi(|{\bf k-\tilde k}| \tau) + 8\tau \Psi(\tilde k\tau)
\Psi'(|{\bf k-\tilde k}| \tau) + 4 \tau^2 \Psi'(\tilde k\tau)\Psi'(|{\bf k-\tilde k}| \tau).
\end{equation}
In Eqs. (\ref{Phktau}, \ref{calF}, \ref{f}) the following notations are used. ${\cal P}_\Psi(k)$ is
the power spectrum of the Bardeen potential, defined at some moment of time
$\tau=\tau_i'$ near the beginning of the RD stage
(by definition, it is the primordial spectrum),
\begin{equation} \label{Ppsilr}
\langle \Psi_{\bf k} \Psi_{\bf k'} \rangle = \frac{2\pi^2}{k^3} \delta^3({\bf k}+{\bf k'}) {\cal P}_\Psi(k),
\end{equation}
$\Psi_{\bf k}$ is the Fourier component of $\Psi$,
\begin{equation}
\Psi({\bf x}) = \frac{1}{(2\pi)^{3/2}} \int d^3 {\bf k} \Psi_{\bf k} e^{i{\bf k}\cdot{\bf x}},
\end{equation}
$\mu = {\bf k \cdot \tilde k} / (k \tilde k)$ is the cosine of the angle between the vectors
${\bf k}$ and ${\bf \tilde k} $. The power spectrum of GWs is defined by the standard expression
\begin{equation}
\langle h_{\bf k}(\tau) h_{\bf k'}(\tau) \rangle = \frac{1}{2} \frac{2\pi^2}{k^3}
\delta^3({\bf k}+{\bf k'}) {\cal P}_h(k, \tau),
\end{equation}
where $h_{\bf k}(\tau)$ is the Fourier component of the tensor metric perturbation,
\begin{equation}
h_{ij}(x, \tau) = \int \frac{d^3 {\bf k}}{(2\pi)^{3/2}} e^{i{\bf k}\cdot{\bf x}}
\left[ h_{\bf k}(\tau) e_{ij}({\bf k}) + \bar h_{\bf k}(\tau) \bar e_{ij}({\bf k})\right],
\end{equation}
$e_{ij}({\bf k})$ and $\bar e_{ij}({\bf k})$ are two polarization tensors corresponding to the
wave number ${\bf k}$.

It had been shown in works of previous authors \cite{Ananda:2006af, Baumann:2007zm}
that the right-hand side of this equation is expressed through the correlator containing the product of four $\Psi_k$
functions, i.e., ${\cal P}_h$ is expressed through the product of two
${\cal P}_\Psi$ spectra, see Eq. (18). So, in this case, like in the case of Eq. (\ref{Ppsilr}),
the average is also over fluctuations of the gravitational potential.

The evolution equation for the GW amplitude is
\begin{equation} \label{h-evol}
h_{\bf k}'' + 2 {\cal H} h_{\bf k}' + k^2 h_{\bf k} = S({\bf k}, \tau),
\end{equation}
where the source term is
\begin{equation}
S({\bf k}, \tau) = \int d^3\tilde {\bf k} \; \tilde k^2(1-\mu^2) \; f({\bf k},\tilde {\bf k}, \tau)
\Psi_{ {\bf k}-\tilde {\bf k}} \Psi_{ \tilde {\bf k} }.
\end{equation}

The function $f$ in Eq. (\ref{f}) contains transfer functions $\Psi(k\tau)$, which are defined by
\begin{equation} \label{PsiTRFUN}
\Psi(k \tau) = \frac{\Psi_k(\tau)}{\Psi_k},
\end{equation}
where $\Psi_k \equiv \Psi_k(\tau_i')$ is the initial (primordial) value of the potential.
During RD epoch, the solution for the Bardeen potential, having the initial condition
$\Psi_k(\tau_i)=0$, where $\tau_i$ is the moment of the end of inflation which is close
to $\tau_i'$ (but $\tau_i<\tau_i'$), is \cite{Lyth:2005ze, Bugaev:2008gw}
\begin{eqnarray}
\label{PsiSol} \Psi_k(\tau, {\cal R}_k) = \frac{2 {\cal R}_k}{x^3} \big[
(x-x_i) \cos(x-x_i) - (1+ x x_i)\sin(x-x_i) \big] \;\;, \;\; x=k\tau/\sqrt{3}.
\end{eqnarray}
Here, the variable ${\cal R}$ is the curvature perturbation on the comoving hypersurfaces
(see, e.g., \cite{Lyth:2005ze}).

The value of the potential at $\tau_i$ is chosen to be zero because $\Psi_k$ is
typically very small during inflation \cite{Lyth:2005ze} and it is a continuous function during
the transition from inflationary to RD stage (we assume, for simplicity, that the reheating
is instant). For matter-dominated (MD) epoch, $\Psi_k(\tau) = {\rm const}$ on all scales.

We have chosen $\tau_i'$
using the condition $\lg(\tau_i'/\tau_i)=0.05$, and due to this the Bardeen
potential $\Psi_k$ at $\tau_i'$ is much smaller than its
asymptotic value $\Psi_k=-(2/3){\cal R}_k$ that is reached in the super-horizon regime ($k\ll aH$)
for $k \ll k_{end} = \tau_i^{-1}$ (this result can be obtained from (\ref{PsiSol}) expanding the sine
and cosine functions).
If we are interested only in such wave numbers ($k\ll k_{end}$),
it is more convenient to define ${\cal P}_\Psi$ in terms of this asymptotic super-horizon value.
To distinguish asymptotic value of the $\Psi$-spectrum from the value at the moment
$\tau_i'$, we will denote it as $\tilde {\cal P}_\Psi$.
For $k \ll k_{end}$, the relation between the two is very simple:
\begin{equation}
\tilde {\cal P}_\Psi =\kappa {\cal P}_\Psi; \;\;
\kappa=\left( \frac{ \frac{2}{3} {\cal R}_k } { \Psi_k } \right)^2,
\label{tildeP}
\end{equation}
and $\Psi_k =\Psi_k(\tau_i', {\cal R}_k)$. For our choice of $\tau_i'$,
$\kappa \approx 11.7$.

The function $g_k(\tau, \tilde\tau)$ in Eq. (\ref{calF}) is the Green function of the Eq. (\ref{h-evol})
which depends on the cosmological epoch. For RD Universe,
\begin{equation}
g_k(\tau, \tilde \tau) = \frac{1}{k} \sin[k(\tau - \tilde\tau)] \;\;\; , \;\;\; \tau < \tau_{\rm eq},
\label{gk-RD}
\end{equation}
and for MD case,
\begin{equation}
g_k(\tau, \tilde \tau) = -\frac{x \tilde x}{k} \left[ j_1(x) y_1(\tilde x) -
j_1(\tilde x) y_1(x) \right] , \;\; x=k\tau , \;\; \tau \ge \tau_{\rm eq}.
\label{gk-MD}
\end{equation}

\subsection{Delta function input power spectrum}

The integral (\ref{Phktau}) is much simplified if we assume an idealized power spectrum with
all power contained in one mode with some wave number $k_0$:
\begin{equation}
{\cal P}_{\Psi}(k) = P_0 \delta\left(\ln \frac{k}{k_0} \right) = P_0 k_0 \delta(k-k_0).
\label{PP0delt}
\end{equation}
Such a case has already been studied in \cite{Ananda:2006af, Saito:2008jc}. The formula for the GW
power spectrum for such an input is obtained from (\ref{Phktau}):
\begin{equation}
{\cal P}_{h}(k) = \frac{ P_0^2 k_0^2 }{k} {\cal F} \left(k,k_0, \mu = \frac{k}{2 k_0}, \tau \right).
\end{equation}
From $\mu=\cos \theta \le 1$, it follows that $k\le 2 k_0$, and GWs in this case
are generated in the frequency interval from $0$ to $2 k_0$.

The corresponding asyptotic value of the coefficient in the delta-spectrum (\ref{PP0delt})
is equal to $\tilde P_0 = \kappa P_0$.

\begin{figure}
\includegraphics[trim = 0 0 0 0, width=0.48 \textwidth]{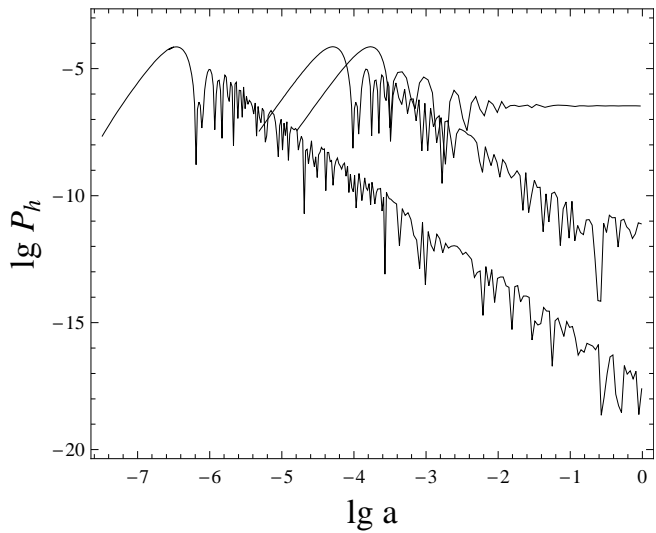} %
\caption{The dependence of ${\cal P}_h(k)$ on the scale factor $a$ for several wave numbers.
As an input, we used here a delta-function power spectrum for ${\cal P}_\Psi$, with
$\tilde P_0=10^{-3}$ for each case.
For curves from top to bottom, $k=k_0=6 k_{eq}, 20 k_{eq}, 3 \times 10^3 k_{eq}$.
} \label{fig-Ph}
\end{figure}

\subsection{GW energy density calculation}

The energy density of GWs per logarithmic interval of $k$ in units of the critical density is given by
\begin{equation} \label{OmegaGW}
\Omega_{\rm gw}(k, \tau) =
\frac{1}{12} \left( \frac{k}{a(\tau) H(\tau)} \right)^2 {\cal P}_h(k, \tau).
\end{equation}
The power spectrum of GWs ${\cal P}_h(k, \tau)$ is obtained from the
formula (\ref{Phktau}). However, for very large
wave numbers $k$ which we are interested in, the direct use of (\ref{Phktau}) will require
numerical integration
for functions having a huge number of oscillations (e.g., for $k \sim 10^{16} {\rm Mpc}^{-1}$
this is about $\sim k \tau_0 \sim 10^{20}$ oscillations). This is hard to do numerically.
Fortunately, we do not have to do integration until the present day. It is enough
to calculate $\Omega_{\rm gw}$ for the moment of time $\tau_{calc} \gg k^{-1}$ at which
the mode is well inside the horizon, and is freely propagating. We can then easily relate
energy densities of GWs at different times with simple calculation, using the fact that
$h_k\sim a^{-1}$ far inside the horizon.
Really, assuming zero source term in (\ref{h-evol}) and changing the variable to $v_k=a h_k$,
we obtain the equation
\begin{equation}
v_k'' + v_k\left[k^2 - a^2 H^2\left(2 - \frac{3}{2}(1+w) \right) \right] = 0,
\end{equation}
where $w=p/\rho$. In sub-horizon regime, when $k \gg aH$, the solution of this equation
is $v_k\sim \cos(k\tau+\varphi)$, and so, ignoring oscillations, $h_k\sim a^{-1}$ .

However, it had been noticed in \cite{Baumann:2007zm} that the propagation of second-order GWs
in sub-horizon regime cannot be always regarded
as free because the source term in the equation (\ref{h-evol}) cannot be neglected in all cases.
We illustrate this point
in Fig. \ref{fig-Ph}, which shows the dependence of ${\cal P}_h(a)$ for several values of $k$,
calculated numerically using Eqs. (\ref{Phktau}, \ref{gk-RD}, \ref{gk-MD}). The input power spectrum
was taken to be of a delta-function form for this example.
It is seen from this figure that
for rather large values of $k$ ($k \gtrsim k_c \approx 100 k_{eq} \approx 1 {\rm Mpc}^{-1}$)
this effect can be neglected,
and a simple relation ${\cal P}_h(a) \sim a^{-2}$ can be used while for smaller $k$ the effect
is important. In the present work we are
interested in GWs with much larger wave numbers than $k_c$, so we can safely use this
relation in our calculations.

\begin{figure}
\includegraphics[trim = 0 0 0 0, width=0.48 \textwidth]{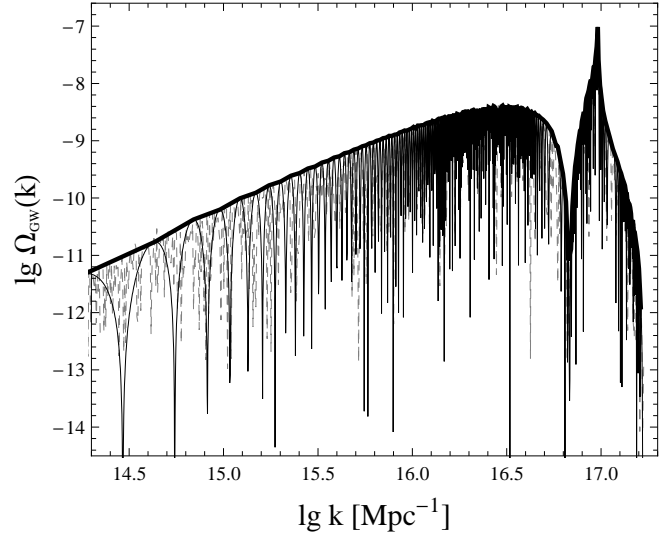} %
\caption{GW spectrum from a delta-function peak in ${\cal P}_\Psi$ ($\tilde P_0=1.2 \times 10^{-3}$,
$k_0=8\times 10^{16}$ Mpc$^{-1}$). Thin solid line - calculation for $\tau_{calc}=10^3 k_0^{-1}$,
thin dashed line - for $\tau_{calc}=(50+N_{\rm rnd})k^{-1}$, thick line is the envelope.} \label{fig-delta}
\end{figure}

During the RD epoch, $aH\sim a^{-1} g_*^{-1/6}$, so
$\Omega_{\rm gw}(k, \tau) \sim (k/aH)^2 {\cal P}_h \sim g_*^{1/3}$, and we can write the relation
for moments of calculation $\tau_{calc}$ and matter-radiation equality $\tau_{eq}$
\begin{equation}
\Omega_{\rm GW}^{eq}(k) = \Omega_{\rm GW}^{calc}(k) \left( \frac{g_{* eq}}{g_{* calc}}\right)^{1/3}.
\end{equation}
After the moment $\tau_{eq}$, $\Omega_{\rm GW}$ is proportional to energy density fraction of
the radiation, which equals $0.5$ at $\tau_{eq}$. So,
\begin{equation}
\Omega_{\rm GW}^0(k) = 2 \Omega_R \times \Omega_{\rm GW}^{eq}(k),
\end{equation}
and, finally,
\begin{equation}
\Omega_{GW}^0(k) =
2 \Omega_R  \left( \frac{g_{* eq}}{g_{* calc}}\right)^{1/3} \times \frac{(k\tau_{\rm calc})^2}{12}
{\cal P}_h(k, \tau_{\rm calc}).
\end{equation}
This formula gives the correct energy density, accurate to the oscillations in it. The exact shape of
the function will, actually, depend on the choice of $\tau_{\rm calc} \gg k^{-1}$, and the
larger $\tau_{\rm calc}$ we take, the more frequent are the oscillations, but the envelope which
we are interested in does not change. In practice,  $\tau_{\rm calc}$ can be either fixed or
dependent on $k$, e.g., for the last case,
\begin{equation}
\tau_{\rm calc} = N_{\rm sub} \cdot k^{-1}, \;\;\;\; N_{\rm sub} \sim 100.
\end{equation}
It proves to be more convenient to use the ``randomized'' value of $N_{\rm sub}$, i.e.,
\begin{equation}
\tau_{\rm calc} = (\tilde N_{\rm sub}+N_{\rm rnd}) \cdot k^{-1},
\end{equation}
where $\tilde N_{\rm sub}$ is constant and $N_{\rm rnd}$ is a random number in the
interval $[0, 2\pi]$ calculated independently for
every $k$. In this case the result of the calculation is a stochastically oscillating function
whose envelope always can be easily found, and it is the envelope that we are interested in.
This argumentation is illustrated in Fig. \ref{fig-delta} where $\Omega_{GW}$ is calculated
for the delta-function power spectrum using two different choices of $\tau_{\rm calc}$
(constant and dependent on $k$). It is seen that the envelope is the same for both cases.

It should be noticed here that for an input spectrum of an ideal $\delta$-function form, the
growth of ${\cal P}_h(k)$ for $k=2k_0/\sqrt{3}$ proceeds logarithmically even when $k\tau \gg 1$
\cite{Ananda:2006af}. It happens because of the ``resonance''
between the functions $g$ and $f$ in integral (\ref{calF}) in the RD epoch. Indeed, omitting the
constant phase shifts,
\begin{equation}
g_k(\tau, \tilde \tau) \sim \sin(k \tau);
\end{equation}
\begin{equation}
f \sim \frac{1}{\tau^n} \sin \left(\frac{\tilde k \tau}{\sqrt{3}} \right)
\sin \left(\frac{|{\bf k - \tilde k}| \tau}{\sqrt{3}} \right)
\sim \frac{1}{\tau^n} \sin^2 \left( \frac{k_0 \tau}{\sqrt{3}} \right),
\end{equation}
and amplification during integration is possible if the condition
\begin{equation}
k \tau = 2 \frac{k_0 \tau}{\sqrt{3}}
\end{equation}
holds, i.e., for the case $k=2k_0/\sqrt{3}$. The width of the resonant peak around this value of $k$
is proportional to $(k\tau)^{-1}$ and its height $\sim \ln (k\tau)$ \cite{Ananda:2006af}, so the power contained
in the peak is small and hardly can be detected. For a realistic spectrum of a
finite width, the ``resonant'' effect still exists, but the amplification continues only
until $\tau \sim 1/\Delta k$, with $\Delta k$ being the characteristic width of the spectrum.

\subsection{Power spectrum with maximum}

It is convenient to use some kind of parametrization to model the realistic peaked power
spectrum of finite width. We use the distribution of the form
\begin{equation}
\label{PRparam} %
\lg {\cal P}_{\cal R} (k) = B + (\lg {\cal P}_{\cal R}^0 - B)
\exp \Big[-\frac{(\lg k/k_0)^2}{2 \Sigma^2} \Big].
\end{equation}
Here, $B \approx -8.6$, ${\cal P}_{\cal R}^0$ characterizes the height of the peak,
$k_0$ is the position of the maximum and $\Sigma$ is the peak's width. Parameters of such
a distribution have been constrained previously \cite{Bugaev:2008gw} from non-observation
of PBHs and products of their Hawking evaporation.

\begin{figure}
\includegraphics[trim = 0 0 0 0, width=0.42 \textwidth]{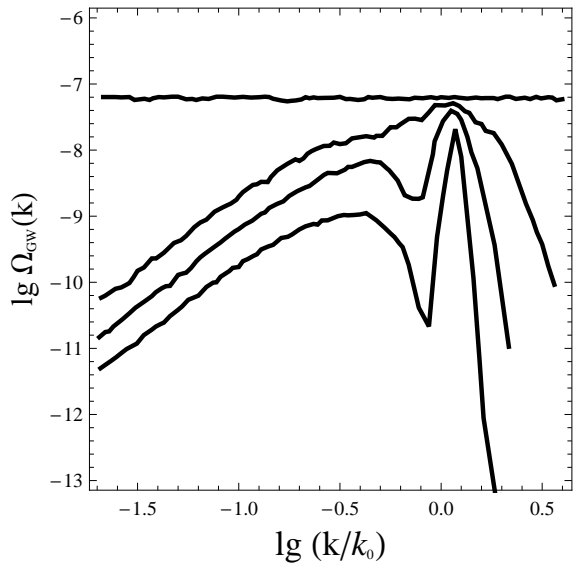} %
\caption{Calculation of $\Omega_{GW}(k)$ at the present epoch for finite width curvature perturbation
power spectra of the form (\ref{PRparam}) (from bottom to top, $\Sigma=0.1, 0.3, 0.8$ and
${\cal P}_{\cal R}^0 = 0.01$; upper curve is for scale-invariant input spectrum with
${\cal P}_{\cal R}(k) = 0.01$). We assumed that $g_*(k_{0})\approx 100$. }
\label{fig-diff-sigma}
\end{figure}

In Fig. \ref{fig-diff-sigma} we show the result of $\Omega_{GW}$ calculation for the finite-width
distribution of the form (\ref{PRparam}). It is seen that for a narrow peak, the distribution
looks like the one produced by a $\delta$-function power spectrum. The shape is smoothing
with the growth of $\Sigma$ and it is scale-invariant for the scale-invariant input. The
value of $\Omega_{GW}$ in this case is proportional to $({\cal P}_{\cal R})^2$ and can be
estimated as
\begin{equation}
\label{OmegaApprox2}
\Omega_{GW}(k>k_c, \tau_0)\cong 0.002 \left( \frac{g_{* eq}}{g_*} \right)^{1/3} \cdot {\cal P}_{\cal R}^2 .
\end{equation}
It follows from this formula that for
obtaining large second-order GW background we need relatively large value of
${\cal P}_{\cal R}$. For instance, for $\Omega_{GW}$ to be of order of $10^{-7}$,
${\cal P}_{\cal R}$ should be of order of $10^{-2}$. In our paper we
are considering examples in which the values of ${\cal P}_{\cal R}$ are
just so large on small scales. The only constraint on ${\cal P}_{\cal R}$ for these scales comes from
PBH studies, and the values can be, in principle, much larger than ${\cal P}_{\cal R}\sim 10^{-9}$
obtained from experiments at large scales.

Note that the shape of the curves shown in Fig. \ref{fig-diff-sigma} does only depend on the
relation $k/k_0$ if other parameters are fixed and $k, k_0 \ll k_{end}$. It can be seen from
(\ref{Phktau}, \ref{calF}): if we make a shift of $k$ (and $\tilde k$) by a constant factor
of $\alpha$ (so, $k\to \alpha k$, $\tilde k\to \alpha \tilde k$, and
${\cal P}_\Psi(k) \to {\cal P}_\Psi(\alpha k)$), and the corresponding shift of calculation time
$\tau_{calc} \to \alpha^{-1} \tau_{calc}$ (so that $k \tau_{calc}$ remains unchanged), then
functions entering the Eq. (\ref{calF}) change in the following proportion:
\begin{equation}
g \sim \frac{1}{k} \sin(k\tau) \sim \frac{1}{\alpha}; \;\; f \sim \Psi^2(k\tau) \sim 1; \;\;
{\cal F} \sim k^3 (d\tau)^2 g^2 f^2 \sim \frac{1}{\alpha}
\label{change-proportion}
\end{equation}
(the change of the integration limits over $\tau$ does not affect the result as soon as
$\tau_0 \ll k^{-1}$). Using (\ref{change-proportion}) and (\ref{Phktau}), we obtain that
\begin{equation}
{\cal P}_h(k, \tau_{calc}) \sim \int dk \int d\mu {\cal P}_\Psi {\cal P}_\Psi {\cal F} \sim \alpha^0,
\label{change-Ph}
\end{equation}
so ${\cal P}_h$, and, correspondingly, $\Omega_{GW}$ do not change at the calculation time, and,
in fact, will have the same values after the shift of $k$ because, as we have seen,
the amplitude of the real GW spectrum does not
depend on $\tau_{calc}$ (as soon as $\tau_{calc} \gg k^{-1}$).

\subsection{Comparison with current experimental data}

The ground-based interferometer LIGO during its fifth science run (S5) have obtained the limit \cite{Nature}
\begin{equation}
\Omega_{GW} < 6.9 \times 10^{-6}. \label{LIGOS5}
\end{equation}
This limit applies to a scale-invariant GW spectrum in the frequency range $41.5-169.25$ Hz.
The previous limit (S4 result) was about an order of magnitude higher \cite{Abbott:2006zx},
\begin{equation}
\Omega_{GW} < 6.5 \times 10^{-5}, \label{LIGOS4}
\end{equation}
for the frequency range $51-150$ Hz. The target
sensitivity of the planned Advanced LIGO experiment is $\Omega_{GW} \sim 10^{-8} - 10^{-9}$
\cite{Abbott:2006zx}.

\begin{figure}
\includegraphics[trim = 0 0 0 0, width=0.51 \textwidth]{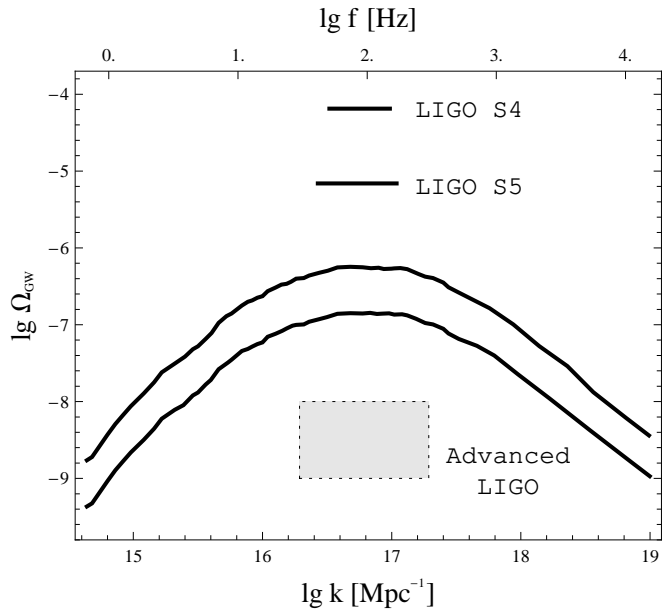} %
\caption{Calculation of $\Omega_{GW}(k)$ at the present epoch for the case $f_0=100$ Hz, $\Sigma=3$,
${\cal P}_{\cal R}^0 = 0.032$ (upper curve) and ${\cal P}_{\cal R}^0 = 0.016$ (lower curve).
Such parameters are maximal allowed from PBH constraints. Also shown are experimental limits
on $\Omega_{GW}$ obtained in the LIGO experiment and bound range expected in the future. }
\label{fig-peak-exp}
\end{figure}

The corresponding horizon mass for the central frequency of LIGO sensitivity range $f\sim 100$ Hz
calculated from (\ref{fMh}) is about $10^{13}$ g. From the other side, it is
known that the intensity of
PBH production in early universe from scalar power distribution of the
form (\ref{PRparam}) in this mass range can be constrained
from studies of photon and neutrino extragalactic diffuse backgrounds
\cite{Bugaev:2008gw}. These constraints lead to corresponding limits of ${\cal P}_{\cal R}^0$.
Particularly, for peak with
horizon mass corresponding to its maximum $M_h^0=10^{13}$ g and width $\Sigma=3$, the constraints
obtained in \cite{Bugaev:2008gw} are ${\cal P}_{\cal R}^0=0.016$ (in case of standard collapse model,
for which critical density contrast leading to PBH formation is $\delta_c=1/3$) or
${\cal P}_{\cal R}^0=0.032$ (for the critical collapse with adopted threshold value $\delta_c=0.45$).
The most optimistic allowed signal of second-order GWs in the LIGO range can thus be estimated
from (\ref{OmegaApprox2}):
\begin{equation}
\Omega_{GW}^{\rm max} \approx 0.002 \times (3/100)^{1/3} \times 0.032^2 \approx 6\times 10^{-7},
\end{equation}
which is an order of magnitude smaller than the current bound (\ref{LIGOS5}), but is reachable for
Advanced LIGO. The results of the full calculation of $\Omega_{GW}$-distribution expected in this case are shown
in Fig. \ref{fig-peak-exp}. It is seen that Advanced LIGO will be capable to reach sensitivity
needed to improve limits on ${\cal P}_{\cal R}$ and PBH abundance in this range of scales.

\section{Running mass model}
\label{sec-RM}

The running mass inflation model was proposed in \cite{Stewart:1996ey, Stewart:1997wg} and further
studied in many papers including \cite{Covi:1998jp, Covi:1998mb, Covi:1998yr, German:1999gi,
Covi:2004tp} and \cite{Bugaev:2008bi}. The model predicts a
rather strong scale dependence of the spectral index, possibly
allowing large values of ${\cal P}_{\cal R}(k)$ at small scales, which can even lead to significant PBH production
and helps to constrain possible model parameters \cite{Leach:2000ea, BugaevD66, Bugaev:2008gw, Alabidi:2009bk}.

\begin{figure}
\includegraphics[trim = 0 0 0 0, width=0.48 \textwidth]{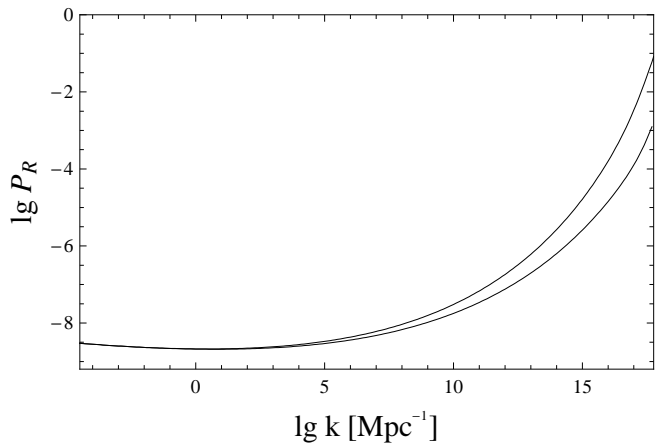} %
\caption{
Power spectrum ${\cal P}_{\cal R}(k)$, calculated for the running mass model, with $T_{RH}=10^{10}$ GeV
and $n_0=0.96$; $n_0'=4.5\times 10^{-3}$ for the upper curve and $n_0'=4.0 \times 10^{-3}$ for the lower curve.}
\label{fig-rm-PR}
\end{figure}

The potential of the running mass model takes into account quantum corrections in the context of the softly
broken global supersymmetry and is given by the formula
\begin{equation}
V = V_0 + \frac{1}{2} m^2(\ln \phi) \phi^2 \;.
\end{equation}
Its shape is determined by parameters $c$ and $s$, which are defined by the relations
\begin{eqnarray}
c \frac{V_0}{M_P^2} &=& - \left. \frac{d m^2}{d \ln \phi} \right|_{\phi=\phi_0} \; , \\
s &=& c \ln \frac{\phi_*}{\phi_0} \; .
\end{eqnarray}
Here, $\phi_*$ is the inflaton field value corresponding to the maximum of the potential, $\phi_0$
is the value at the epoch of horizon exit for the pivot scale $k_0\approx 0.002 h$ Mpc$^{-1}$.
The same parameters determine the behavior of the power spectrum of density perturbations.
The connection of $s$, $c$ with the observable quantities $n_0$ and $n_0'$ (the spectral index
and its running, respectively) is given by
\begin{equation}
n_0-1 \approx 2 (s-c) \;\; , \;\; n_0' \approx 2 s c \;.
\end{equation}
According to analysis of \cite{Bugaev:2008bi}, the possible choice is
\begin{equation}
c \approx 0.06 \;\; ; \;\; s \approx 0.04,
\end{equation}
corresponding to $n_0\approx 0.96$ and $n_0'\approx 0.005$. In the case of positive $s$ and $c$, the
inflaton field decreases during inflation. It means that the slow-roll parameter $\epsilon$ also
decreases:
\begin{equation}
V\sim V_0 \; , \; \epsilon \sim \left( \frac{V'}{V} \right)^2 \sim \phi^2.
\end{equation}
Correspondingly, the behaviors of the power spectra for density perturbations and
gravitational waves are strongly different,
\begin{equation}
{\cal P}_{\cal R} \sim \frac{V_0}{\epsilon} \; , \; {\cal P}_h^{\rm (infl)} \sim V_0 \; .
\end{equation}
As a result, the power spectrum of induced GWs may be quite substantial at small scales
in spite of the fact that ${\cal P}_h^{\rm (infl)}$ is negligibly small. The parameter $s$ connects
the field value $\phi_0$ with the Hubble parameter during inflation and with the
normalization of the CMB power spectrum:
\begin{equation} \label{phi0s}
\phi_0 s = \frac{H_I}{2 \pi  {\cal P}_{\cal R}^{1/2}(k_0) }.
\end{equation}
From a theoretical point of view \cite{Covi:2004tp},
$H_I$ can lie in the wide range of values, depending on the mechanism for supersymmetry breaking,
from $H_I\sim 10^4$ GeV for ``anomaly-mediation'' case to $H_I \sim 10^{-3}$ GeV for
``gauge-mediation'' (we suppose that the inflationary potential is of order of $M_{infl}^4$
where $M_{infl}$ is the scale of supersymmetry breaking during inflation which is approximately equal to
the scale of supersymmetry breaking in the vacuum). Strictly speaking, the form and amplitude of the
power spectrum depend on the value of $H_I$, according to Eq. (\ref{phi0s}). But we will assume,
for simplicity, that the power spectrum behavior is determined mainly by parameters $s$ and $c$,
independently on $H_I$, while the value of $H_I$ determines only the temperature of reheating, $T_{RH}$,
and a value of the comoving scale crossing the horizon at the end of inflation, $k_{end}$.
If the reheating is instant, these values are given by
\begin{eqnarray}
H_I &\cong& \frac{\pi}{ g_*^{1/2} 3\sqrt{10}} \frac{T_{RH}^2}{M_p} \; , \\
k_{end} &\approx& 2.6 \times 10^7 g_*^{1/6} \left( \frac{T_{RH}}{1 \; {\rm GeV}} \right) {\rm Mpc}^{-1} \;.
\label{kend}
\end{eqnarray}

The tensor power spectrum generated during inflation and the amplitude of the
 inflationary GWB are determined by the energy scale of inflation
\begin{equation}
\label{PhSR}
{\cal P}_h^{\rm (infl)}(k) \approx \left. \frac{16 H^2}{\pi m_{Pl}^2} \right|_{k=aH},
\end{equation}
and, because $H \approx H_I$, we have ${\cal P}_h^{\rm (infl)} \lesssim 10^{-30}$
even for the largest possible value of $H_I$.
This is too small to be detected even for the most sensitive experiments proposed.

\begin{figure}
\includegraphics[trim = 0 0 0 0, width=0.51 \textwidth]{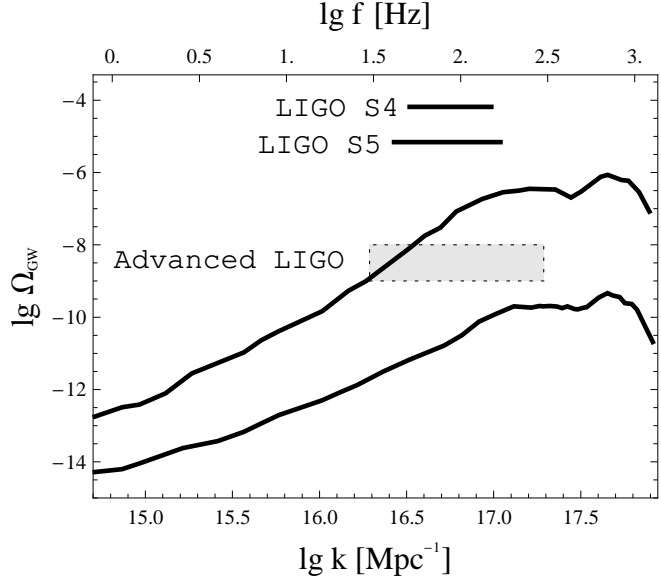} %
\caption{Calculation of $\Omega_{GW}(k)$ for the power spectrum of curvature perturbations
expected in the running mass model (see Fig. \ref{fig-rm-PR}). We used the following set of parameters for the
calculation: $T_{RH}=10^{10}$ GeV, $n_0=0.96$ (both curves); $n_0'=0.0045$ for the upper
curve and $n_0'=0.004$ for the lower one. Also shown are the experimental limits
on $\Omega_{GW}$ obtained in the LIGO experiment and bound range expected in the future.}
\label{fig-rm-gw}
\end{figure}

The maximum values of frequencies of the GW spectrum depend on the moment of time of
phase transition to RD stage from the preceding stage (see,
e.g., \cite{Allen:1996vm}). For more general discussion about the form of the relic GW
spectrum see the recent review \cite{Grishchuk:2007uz}. In particular, in
inflationary models, the maximum value of the frequency depends
on the reheating temperature, see, e.g., Eq. (\ref{kend}).
Our model with running mass potential predicts large values of the curvature perturbation
spectrum just near $k_{end}$, i.e., the position of the maximum of second-order GW background
also depends on the reheating temperature.

The numerically calculated power spectrum ${\cal P}_{\cal R}(k)$ in the running mass model
is shown in Fig. \ref{fig-rm-PR} for two sets of parameters.
Parameters $n_0$, $n_0'$ of the spectra are chosen with taking into account the PBH constraints obtained
in the previous work of authors \cite{Bugaev:2008gw}. It is seen from the figure, that the maximum values
of the power spectrum (at large $k$) are very sensitive to the value of the spectral index running.
The upper curve in Fig. \ref{fig-rm-PR} represents the spectrum with largest possible $n_0'$
for other parameters being fixed. Larger values of spectral index running will cause too
large curvature perturbation spectrum values - in this case, too many PBHs will be produced
with mass $\sim 3\times 10^{11}$ g.

\begin{figure}
\includegraphics[trim = 0 0 0 0, width=0.51 \textwidth]{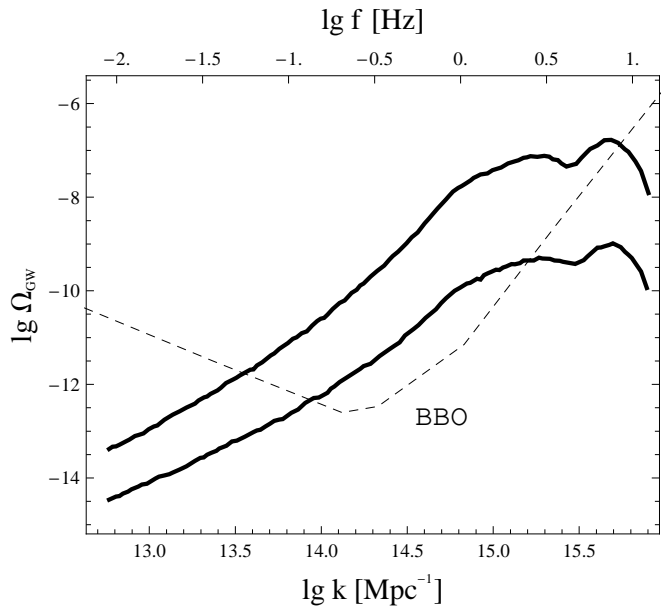} %
\caption{Calculation of $\Omega_{GW}(k)$ for the
running mass model. For this Figure, we used the following set of parameters in the
calculation: $T_{RH}=10^{8}$ GeV, $n_0=0.96$ (both curves); $n_0'=0.0054$ for the upper
curve and $n_0'=0.005$ for the lower one. We also show the expected sensitivity
curve for the proposed BBO experiment \cite{Smith:2008pf}.}
\label{fig-rm-gw-bbo}
\end{figure}

The spectra of second-order GWs corresponding to the scalar spectra of the running mass model
are shown in Figs. \ref{fig-rm-gw}, \ref{fig-rm-gw-bbo}. As in the previous
case of a peaked-power spectrum (Fig. \ref{fig-peak-exp}), it is seen
from these figures that scalar perturbation power spectrum generated in the running mass model
can be a source of significant amount of GWs, detectable in experiments studying the frequency region
$\sim 10^{-1} - 10^{3}\;$Hz.

The results shown in Figs. \ref{fig-rm-gw}, \ref{fig-rm-gw-bbo} correspond to different values of
$k_{end}$ and $T_{RH}$  (i.e., to different energy scales of inflation). It was shown in
\cite{Bugaev:2008gw} that PBH constraints depend on a proposed value of $T_{RH}$: the maximum
running allowed is $n_0'=0.0045$ for $T_{RH}=10^{10}$ GeV and $n_0'=0.0054$ for $T_{RH}=10^{8}$ GeV
(for $n_0=0.96$ in both cases). The difference in PBH constraints originates from the connection between
comoving scale and horizon mass (see Eq. \ref{kkeq}): $k\sim k_{end}\approx 6\times 10^{15}\;$Mpc$^{-1}$
corresponds to  $M_{BH} \sim M_h \approx 3 \times 10^{15}\;$g, while
$k\sim k_{end}\approx 6\times 10^{17}\;$Mpc$^{-1}$ corresponds to  $M_{BH}\sim M_h \approx 3 \times 10^{11}\;$g.
PBH constraints in the region of $M_h\sim 10^{15}\;$g follow from inspection of extragalactic photon
background data while PBH constraints near $M_h\sim 10^{11}\;$g follow from available limits
on the intensity of extragalactic neutrino background \cite{Bugaev:2008gw}.

\section{Conclusions}
\label{sec-concl}

The main new results of the paper are shown in Figs. \ref{fig-diff-sigma}, \ref{fig-peak-exp} and
\ref{fig-rm-gw}, \ref{fig-rm-gw-bbo}.

1. The dependence of a form of
the $\Omega_{GW}(k)$-curve on the width of the peak of the
scalar power spectrum is carefully studied (Fig. \ref{fig-diff-sigma}).
It is shown that this curve has a characteristic ``double-peak'' form; the height
of the larger maximum weakly depends on the width of the proposed ${\cal P}_{\cal R}(k)$
distribution (see Eq. (\ref{PRparam})). It is shown also that the $\Omega_{GW}(k)$-function
depends, in fact, on the ratio $k/k_0$ where $k_0$ is the position of the maximum in
${\cal P}_{\cal R}(k)$.

2. The distribution $\Omega_{GW}(k)$ for the induced GWB is calculated for two different
cases: for the scalar spectrum ${\cal P}_{\cal R}(k)$ with a peak (Eq. (\ref{PRparam})) and
for the scalar power spectrum predicted by the running mass model. It is seen from the
resulting figures that the behavior of $\Omega_{GW}(k)$-curves near the maximum in both cases
is rather similar.

3. It is shown that maximum values of $\Omega_{GW}(k)$-distributions for the induced GWB
can be constrained by PBH searches in the wide region of the comoving scales,
$10^{14} \lesssim k \lesssim 10^{18}\;$Mpc$^{-1}$ (the upper curves in Figs \ref{fig-peak-exp},
\ref{fig-rm-gw}, \ref{fig-rm-gw-bbo} are maximum ones in this sense). It is shown also that, up
to now, the PBH constraints are more strong than the available limits from LIGO
(see Fig. \ref{fig-peak-exp} and \ref{fig-rm-gw}). In particular, it follows from
Fig. \ref{fig-peak-exp} that the maximum value of $\Omega_{GW}$ at frequency $f\sim 100\;$Hz
is equal to $6\times 10^{-7}$ (supposing that the scalar power spectrum has the peak in this region).

Finally, it follows from Fig. \ref{fig-peak-exp} that, in not very distant future, when
the limits for $\Omega_{GW}$ from LIGO experiment will reach the level $\sim 10^{-8}$ and
lower, their data will give stronger constraints on amplitudes of scalar power spectra then
today's constraints following from PBH studies. It means, that the future experiments with
LIGO detector, limiting the power spectrum of primordial scalar perturbations, will give the
new constraints for PBH production in the early universe. The interval of frequencies of
ground-based detectors is about $\sim 10$ Hz to $\sim$ few kHz, and the corresponding
interval of PBH masses is from  $\sim 10^{11}$ to $\sim 10^{15}$ g.

%\pagebreak

In this paper we have made an accent on second-order GWs with frequencies $\sim$ mHz - kHz,
which will be probed by space-based and ground-based laser interferometer experiments.
The same effects (particularly, rather large values of induced $\Omega_{GW}$) can appear
in other frequency regions, for which ${\cal P}_{\cal R}$ is not probed by observations
($f \gtrsim 10^{-14}\;$Hz), e.g., for $f\gtrsim 10^5\;$Hz, where high-frequency GW detectors
operate, or at $f\sim 10^{-8}\;$Hz, where $\Omega_{GW}$ is constrained by pulsar timing
data (in fact, these data already allow to put limits on PBH abundance \cite{Saito:2008jc}
for PBHs with rather large masses). In the band probed by CMB measurements, the
effects from second-order GWs are carefully studied in \cite{Baumann:2007zm} and shown to be small.

\paragraph*{Acknowledgments.}
The work was supported by Federal Agency for Science and Innovation under state
contract 02.740.11.5092.

\end{document}